\documentstyle[11pt,epsfig]{article}
\newcounter{subequation}[equation]

\makeatletter

\def\thesubequation{\theequation\@alph\c@subequation}
\def\@subeqnnum{{\rm (\thesubequation)}}
\def\slabel#1{\@bsphack\if@filesw {\let\thepage\relax
   \xdef\@gtempa{\write\@auxout{\string
      \newlabel{#1}{{\thesubequation}{\thepage}}}}}\@gtempa
   \if@nobreak \ifvmode\nobreak\fi\fi\fi\@esphack}
\def\subeqnarray{\stepcounter{equation}
\let\@currentlabel=\theequation\global\c@subequation\@ne
\global\@eqnswtrue
\global\@eqcnt\z@\tabskip\@centering\let\\=\@subeqncr
$$\halign to \displaywidth\bgroup\@eqnsel\hskip\@centering
  $\displaystyle\tabskip\z@{##}$&\global\@eqcnt\@ne
  \hskip 2\arraycolsep \hfil${##}$\hfil
  &\global\@eqcnt\tw@ \hskip 2\arraycolsep
  $\displaystyle\tabskip\z@{##}$\hfil
   \tabskip\@centering&\llap{##}\tabskip\z@\cr}
\def\endsubeqnarray{\@@subeqncr\egroup
                     $$\global\@ignoretrue}
\def\@subeqncr{{\ifnum0=`}\fi\@ifstar{\global\@eqpen\@M
    \@ysubeqncr}{\global\@eqpen\interdisplaylinepenalty \@ysubeqncr}}
\def\@ysubeqncr{\@ifnextchar [{\@xsubeqncr}{\@xsubeqncr[\z@]}}
\def\@xsubeqncr[#1]{\ifnum0=`{\fi}\@@subeqncr
   \noalign{\penalty\@eqpen\vskip\jot\vskip #1\relax}}
\def\@@subeqncr{\let\@tempa\relax
    \ifcase\@eqcnt \def\@tempa{& & &}\or \def\@tempa{& &}
      \else \def\@tempa{&}\fi
     \@tempa \if@eqnsw\@subeqnnum\refstepcounter{subequation}\fi
     \global\@eqnswtrue\global\@eqcnt\z@\cr}
\let\@ssubeqncr=\@subeqncr
\@namedef{subeqnarray*}{\def\@subeqncr{\nonumber\@ssubeqncr}\subeqnarray}
\@namedef{endsubeqnarray*}{\global\advance\c@equation\m@ne%
                           \nonumber\endsubeqnarray}

\makeatletter
\@addtoreset{equation}{section}
\makeatother
\renewcommand{\theequation}{\thesection.\arabic{equation}}

\def\dalemb#1#2{{\vbox{\hrule height .#2pt
        \hbox{\vrule width.#2pt height#1pt \kern#1pt
                \vrule width.#2pt}
        \hrule height.#2pt}}}
\def\square{\mathord{\dalemb{6.8}{7}\hbox{\hskip1pt}}}

\def\half{{\textstyle{1\over2}}}
\let\a=\alpha \let\b=\beta  \let\d=\delta \let\e=\epsilon
  \let\q=\theta  \let\k=\kappa
\let\l=\lambda \let\m=\mu \let\n=\nu  \let\p=\pi \let\r=\rho
\let\s=\sigma \let\t=\tau  \let\f=\phi \let\c=\chi 
       \let\D=\Delta  \let\L=\Lambda
 \let\P=\Pi   \let\F=\Phi 
\let\C=\Chi 
\let\la=\label  
  
\def\nn{\nonumber} \def\bd{\begin{document}} \def\ed{\end{document}}
\def\ds{\documentstyle} \let\fr=\frac \let\bl=\bigl \let\br=\bigr
\let\Br=\Bigr \let\Bl=\Bigl
\let\bm=\bibitem
\let\na=\nabla
\let\pa=\partial \let\ov=\overline
\def\ie{{\it i.e.\ }}
\newcommand{\be}{\begin{equation}}
\newcommand{\ee}{\end{equation}}
\def\ba{\begin{array}}
\def\ea{\end{array}}
\def\ft#1#2{{\textstyle{{\scriptstyle #1}\over {\scriptstyle #2}}}}
\def\fft#1#2{{#1 \over #2}}
\def\del{\partial}
\def\sst#1{{\scriptscriptstyle #1}}
\def\oneone{\rlap 1\mkern4mu{\rm l}}
\def\e7{E_{7(+7)}}
\def\td{\tilde}
\def\wtd{\widetilde}
\def\im{{\rm i}}
\def\bog{Bogomol'nyi\ }
\def\q{{\tilde q}}
\def\hast{{\hat\ast}}
\def\0{{\sst{(0)}}}
\def\1{{\sst{(1)}}}
\def\2{{\sst{(2)}}}
\def\3{{\sst{(3)}}}
\def\4{{\sst{(4)}}}
\def\5{{\sst{(5)}}}
\def\6{{\sst{(6)}}}
\def\7{{\sst{(7)}}}
\def\8{{\sst{(8)}}}
\def\n{{\sst{(n)}}}
\def\oo{{\"o}}
\def\hA{\hat{\cal A}}
\def\ns{{\sst {\rm NS}}}
\def\rr{{\sst {\rm RR}}}
\def\tH{{\widetilde H}}
\def\tB{{\widetilde B}}
\def\cA{{\cal A}}
\def\cF{{\cal F}}
\def\tF{{\wtd F}}
\def\Z{\rlap{\sf Z}\mkern3mu{\sf Z}}
\def\ep{{\epsilon}}
\def\IIA{{\rm IIA}}
\def\IIB{{\rm IIB}}
\def\ads{{\rm AdS}}
\def\R{\rlap{\rm I}\mkern3mu{\rm R}}
\def\mapright#1{\smash{\mathop{-\!\!\!-\!\!\!-\!\!\!-\!\!\!-\!\!\!
             \longrightarrow}\limits^{#1}}}

\def\ba {\begin{eqnarray}}
\def\ea {\end{eqnarray}}
\def\nn {\nonumber}
\def\half{{1\over2}}
\def\a  {\alpha}
\def\b  {\beta}
\def\c  {\gamma}
\def\C  {\Gamma}
\def\d  {\delta}
\def\D  {\Delta}
\def\e  {\epsilon}
\def\F  {\Phi}
\def\k  {\kappa}
\def\l  {\lambda}
\def\L  {\Lambda}
\def\m  {\mu}
\def\n  {\nu}
\def\o  {\omega}
\def\O  {\Omega}
\def\p  {\pi}
\def\P  {\Pi}
\def\r  {\rho}
\def\th {\theta}
\def\s {\sigma}
\def\t  {\tau}
\def\la {\label}
\def\le {\left}
\def\ri {\right}
\def\pa {\partial}
\def\f {\frac}
\def\sq {\sqrt}
\def\no {\noindent}
\def\bi {\begin{itemize}}
\def\ei {\end{itemize}}
\def\np {\newpage}
\def\ra {\rangle}
\def\vs {\vspace}
\def\llra {\Longleftrightarrow}
\def\veck {\vec{k}_\perp}
\def\vecm {|\vec{k}_\perp|}
\def\pl {{\cal P}}
\def\bfy{{\bf y}}
\def\bfk{{\bf k}}
\def\Ei{{\hbox{Ei}}}
\def\Ci{{\hbox{Ci}}}
\def\Si{{\hbox{Si}}}

\newcommand{\ho}[1]{$\, ^{#1}$}
\newcommand{\hoch}[1]{$\, ^{#1}$}
\newcommand{\bea}{\begin{eqnarray}}
\newcommand{\eea}{\end{eqnarray}}
\newcommand{\lra}{\longrightarrow}
\newcommand{\Lra}{\Leftrightarrow}
\newcommand{\aap}{\alpha^\prime}
\newcommand{\bp}{\tilde \beta^\prime}
\newcommand{\tr}{{\rm tr} }
\newcommand{\Tr}{{\rm Tr} }
\newcommand{\NP}{Nucl. Phys. }

\newcommand{\brussels}{\it Physique Th\'eorique et Math\'ematique,
Universit\'e Libre de Bruxelles,\\ Campus Plaine C.P. 231, B-1050
Bruxelles, Belgium}

\newcommand{\auth}{J.F. V\'{a}zquez-Poritz}

\begin{document}
\begin{flushright}
ULB-TH/02-25\\
September 2002\\
\hfill{\bf hep-th/0209194}\\
\end{flushright}

\begin{center}

{\large {\bf Phases of Braneworlds, Spinning D3-branes and\\ 
Strongly-Coupled 
Gauge Theories}}

\vspace{20pt}

\auth

\vspace{10pt}
\hoch{}\brussels\\

\vspace{30pt}

\underline{ABSTRACT}
\end{center}

A spinning nonextremal D3-brane undergoes a phase transition to a
naked singularity which, from the braneworld point of view, corresponds to
the apparent graviton speed passing from subluminal to superluminal. 
We investigate this phase transition from the dual perspectives of
braneworld scenarios and holography. We discuss the relevance of the
thermodynamic stability domains of a spinning D3-brane to the physics of
braneworld scenarios. We also describe various gravitational
Lorentz violations which arise from static D3-branes.

{\vfill\leftline{}\vfill\vskip 10pt \footnoterule {\footnotesize
jvazquez@ulb.ac.be

\vskip  -12pt} \vskip   14pt
}

\pagebreak
\setcounter{page}{1}

\section{Introduction}

Many of the most interesting solutions of General Relativity contain
spacetime singularities, such as black holes and cosmological
solutions. Brane solutions of M-theory contain classically
singular spacetimes, many of which are classically exact solutions.
It is believed that, close to singularities, classical gravity no longer
suffices. This remains persistent motivation in the search for a quantum
theory of gravitation. A key question is: do spacetime singularities even
exist at all within the quantum gravitational regime of M-theory (see
{\cite{nat} for a recent review)? 

On the other hand, some singularities can actually be resolved at the
classical level, such as by oxidizing to higher dimensions \cite{townsend}, 
T-dualizing to regular solution, or adding flux corresponding to wrapped
branes \cite{kleb,resolve}. Other singularities have been deemed
``unphysical" in the first place, which often implies that the gravitational 
force becomes repulsive near the singularity\footnote{An example of such a 
singular spacetime is the negative-mass Schwarzschild black hole. States
with arbitrarily negative energy would be allowed and the vacuum, and therefore
the theory, would not be stable.}. Thus, rather than being 
``resolved," such singularities are ``excised" \cite{myers,gub,mald}. One
conjecture which prohibits unphysical singularities is the cosmic
censorship. In its weak form \cite{penrose1}, the cosmic censorship
conjecture states that singularities are cloaked by event horizons and, in
its strong form \cite{penrose2}, that even observers falling into black
holes cannot observe a singularity.

M-theory implies the existence of extra dimensions, which are hidden
either via compactification or the localization of gravity on a brane
\cite{randall2,randall}. In this paper, we focus on the second
possibility. In particular, if all observers are confined to a braneworld
then it need not be necessary to prohibit a naked singularity in the bulk,
so long as there are appropriate boundary conditions for the gravitons at
the singularity \cite{gub}. That is, no conserved quantities should be
lost through the singularity. In fact, a singular bulk geometry might
have desirable effects, such as helping to solve the cosmological constant 
problem \cite{verlinde,arkani,kachru}. 

In addition, for some braneworld scenarios with a naked singularity in
the bulk, the graviton speed increases further in the bulk
\cite{Kalbermann,Chung,Ishihara,dan,csaki,csaki2,csaki3}. In this case,
gravitons may traverse distances on the braneworld by bending in the extra
dimension. Since photons remain on the braneworld, gravitons may travel at
an average speed that is greater than the speed of light on the
braneworld. This may offer a non-inflationary solution to the cosmological
horizon problem. On the other hand, if the singularity is hidden within
the event horizon of a black hole in the bulk, then the graviton speed
decreases in the bulk. 

Extremal black holes in an asymptotic $AdS_5$ spacetime were derived in
\cite{behrndt1}, and their nonextremal generalization in \cite{behrndt2}.
Spinning branes were first constructed in \cite{youm}. There is a
one-to-one correspondence between spinning branes and R-charged AdS-black
holes \cite{cvet1} and the non-linear spherical dimensional reduction of
supergravities relating various spinning branes with AdS-black holes was
given in \cite{duff}. In particular, lifting five-dimensional AdS-black
holes to ten dimensions as the near-horizon region of spinning D3-branes 
enables us to explore the proposed complementarity \cite{liu} of
holography \cite{maldacena} and braneworld scenarios
\cite{randall2,randall} in terms of phase transitions. At the
same time, a study of both the gravitational and gauge sectors
within this context may lead to a more complete understanding of the
nature of these phase transitions. The holographic interpretation of
asymmetrically warped spacetimes, for which the bulk singularity is hidden
behind an event horizon, was discussed in \cite{crem}.

This paper is organized as follows. In the next section, we briefly
discuss five-dimensional AdS-black holes/singularities, which arise from
the dimensional reduction of spinning D3-branes in ten dimensions. The
local speed of gravitational propagation is given. In section 3, we
consider braneworlds from static distributions of D3-branes. Braneworlds
arising from extremal flat D3-brane distributions are briefly
reviewed. Next, it is pointed out that apparently-superluminal gravitational
propagation arises only for branes with spherical spatial surfaces
($k=1$), for which we calculate exactly the apparent graviton speed. For
flat nonextremal branes, gravitational Lorentz violations
result in a massive graviton on the brane. In section 4, we study how
certain phase transitions are manifest in both the braneworld model and
the dual gauge theory. We also consider the stability domains for a
spinning D3-brane and the ramifications this has to the resulting
braneworld models. We summarize our results in section 5.

\section{Spinning D3-branes and the speed of gravitons}

In the near-horizon decoupling limit, a $p$-brane spinning in transverse
directions dimensionally reduces to a charged black hole in a domain-wall
background \cite{cvet1,duff}. The angular momenta of the $p$-brane are
interpreted as the black hole electric charges under a subgroup of the
R-symmetry group. For the particular case of the spinning D3-brane, the
(nontrivial) $S^5$-reduction of type IIB supergravity yields an AdS-black
hole in five-dimensions, whose metric is \cite{behrndt1,behrndt2}
\be
ds_5^2=-(H_1 H_2 H_3)^{-2/3}f\, dt^2+(H_1 H_2 H_3)^{1/3}(f^{-1}dr^2+
g^2 r^2 dx_j^2),\label{metric1}
\ee
where
\be
f=k-\frac{\mu}{r^2}+g^2 r^2 (H_1 H_2 H_3), \quad H_i=1+\frac{R_i^2}{k\, 
r^2}, \label{f}
\ee
and $i,j=1,2,3$.

In (five) ten dimensions, the quantities $1/g$, $R_i$ and $\mu $ are
interpreted as the (AdS radius) D3-brane charge, (AdS-black hole
charge) angular momenta and (AdS-black hole mass) energy above 
extremality, respectively. The extremal limit $\mu=0$ corresponds to a
static distribution of D3-branes. $k=1,0$ or $-1$, corresponding to the
foliating surfaces with metric $dx_j^2$ being $S^3$, $T^3$ or $H^3$. For
the case $k=0$, one does the rescaling $R_i^2/k \longrightarrow R_i^2$
before sending $k$ to zero. 

The largest value of $r$ for which $f=0$ corresponds to an event
horizon. For large enough $R_i$, there is no event horizon-- i.e., there
is a naked singularity. The values of $R_i$ and $\mu$ for which this phase
transition ocurrs can be calculated either by setting the horizon radius
to zero or by setting the temperature of the system to zero. For a
distribution of branes, the geometry is singular near the
distribution when $\mu=0$. This will be the first class of singular
braneworld geometries which we will study, in the next section. For a
single nonzero $R_i$, there is an event horizon for $\mu \neq 0$. For
a two nonzero and equal $R_i$, the phase transition to a singular geometry
takes place for $g^2 R^4=\mu$. The braneworld scenario from such a
geometry is discussed in section 4.1. Finally, for three equal $R_i$, the
phase transition to a singular geometry is at $g^2 R^4=\frac{4}{27}\mu$,
as discussed in section 4.2.

With a coordinate transformation, the metric (\ref{metric1}) can be
expressed as
\be
ds_5^2=(H_1 H_2 H_3)^{1/3}g^2 r^2 \Big( -\frac{f}{g^2 r^2(H_1 H_2
H_3)}dt^2+dx_i^2 +dz^2 \Big), \label{metric}
\ee
where 
\be
dz=dr/(r \sqrt{f}). \label{coord}
\ee
In general, $r(z)$ is the amplitude of a Jacobi elliptic 
function. However, frequently a less complicated coordinate transformation
suffices for our purposes.  

The local speed of gravitational propagation is given by
\be
v(z)=\sqrt{\frac{f}{ g^2 r^2 H_1 H_2 H_3}}. \label{v}
\ee
A braneworld may lie at $z=0$. While Standard Model particles are
restricted to live on the braneworld, gravitons may propagate in the bulk
dimension $z$. In particular, if the speed $v(z)$ increases away from
$z=0$, then graviton geodesics bend into the bulk, taking the path of
least-time, and thereby traverse distances on our braneworld faster than
photons.

\section{Braneworlds from static D3-branes}

\subsection{Extremal flat D3-brane distributions}

As can easily be seen from (\ref{v}), for a braneworld scenario arising
from the near-horizon region of a flat ($k=0$) and extremal ($\mu=0$) 
D3-brane, the speed of gravitons in the bulk is the same as the speed of
light on the braneworld. For completion, we will first briefly review
such examples. In the extremal limit, the angular momenta $R_i$ become
distribution parameters for a static solution. The original RS2 model
\cite{randall} can be dimensionally-lifted to a stack of D3-branes 
\cite{cvetic} (all $R_i=0$). There is no mass gap between the localized
massless graviton and the continuous spectrum of massive modes in the
bulk.

Nonzero $R_i$ corresponds to a distribution of D3-branes. From the
holographic viewpoint, separating some of the D3-branes in the transverse
space moves the dual ${\cal N}=4$ super Yang-Mills theory onto the Coulomb
branch \cite{douglas,bilal,wu,tseytlin,larsen}, for which certain scalar 
fields have nonzero expectation values. For a braneworld scenario arising
from D3-branes distributed uniformly over a disc or three-sphere ($R_1 \ne
0$, $R_2,R_3=0$ or $R_1=R_2 \ne 0, R_3=0$, respectively) \cite{sfetsos},
there is a mass gap in the graviton spectrum. In the latter case, the
massive graviton spectrum is discrete. The braneworld scenario arising
from D3-branes distributed uniformly over a five-sphere (all $R_i$
equal) does not appear to have been previously studied. The geometry is
AdS for $r>R_i$ and Minkowski for $r<R_i$. The resulting braneworld
scenario is similar to the previous case, in that there is a mass gap and
the massive graviton spectrum is discrete. These are our first examples of
braneworld scenarios which arise from singular geometries, since the
curvature blows up close to the brane distributions. Although corrections
due to higher-derivative terms in the action become important, the large
curvature region is small, and so it is not expected that the qualitative
picture is altered \cite{sfetsos,freed,brand}.

\subsection{Uniform disc of extremal D3-branes (with unspecified $k$)}

For D3-branes distributed uniformly over a disc, with $k$ unspecified,
(\ref{metric}) and (\ref{coord}) yield
\be
ds_5^2=\frac{1}{{\rm sinh}^2 (1+g|z|)} [1+\gamma^2 {\rm sinh}^2
(1+g|z|)]^{1/3}\Big( \frac{{-\rm cosh}^2 (1+g|z|)}{1+\gamma^2 {\rm
sinh}^2 (1+g|z|)} dt^2+dx_i^2+dz^2\Big),
\label{background}
\ee
where
\be
\gamma \equiv \frac{gR}{\sqrt{k+(gR)^2}}.
\ee

We take $dx_i^2=g^{-2}d\Omega_3^2$, so that $g\rightarrow 0$ for the case
$k=0$. This case corresponds to flat D3-branes uniformly distributed over a 
disc, for which $\gamma=1$. The resulting braneworld scenario
\cite{sfetsos} has been discussed briefly in the previous section. $\gamma=0$ 
corresponds to a D3-brane with $k=1$, and the resulting braneworld
scenario exhibits a graviton speed which increases further in the bulk and
leads to an apparent ``superluminal" propagation of gravity
\cite{vazquez}. 

$\gamma$ is a useful parameter for interpolating between the two
aforementioned cases, by varying $gR$ relative to $k$. As we will see, the
graviton wave equation and spectrum is independent of $\gamma$, so that
the nature of gravitational Lorentz violations can be studied
independently of the graviton spectrum and the localization of the
massless graviton. 

We shall now show that there is a massless graviton mode localized on the
braneworld. The fluctuations of the five-dimensional graviton satisfy the
equation for a minimally-coupled scalar field given by
\be
\partial_M \sqrt{-g}g^{MN}\partial_N \Phi=0. \label{minimal}
\ee
Taking $\Phi=\phi(z) M(t,x_i)$, the radial wave equation for the
background metric (\ref{background}) is
\be -\frac{{\rm sinh}^3(1+g|z|)}{{\rm cosh}(1+g|z|)}\partial_z
\frac{{\rm cosh}(1+g|z|)}{{\rm sinh}^3(1+g|z|)}\partial_z \phi=m^2
\phi, \label{wave} \ee
where $m$ is defined by
\be
\square_{(4)} M(t,x_i)=m^2 M(t,x_i),
\ee
where $\square_{(4)}$ is the four-dimensional Laplacian.

With the wave function transformation
\be 
\phi=\Big( \frac{{\rm cosh}(1+g|z|)}{{\rm sinh}^3(1+g|z|)}
\Big)^{-1/2} \psi, \label{eom}
\ee
the wave equation (\ref{wave}) can be expressed in Schr\"{o}dinger
form,
\be 
-\partial_z^2 \psi +V(z)\psi=m^2 \psi, 
\ee
with
\be 
V(z)=\frac{4\, {\rm sinh}^4 (1+g|z|)+20\, {\rm sinh}^2
(1+g|z|)+15} {{\rm sinh}^2 (2(1+g|z|))}g^2-\alpha\, g\, \delta
(z), 
\ee
where $\alpha=2(2{\rm sinh}^2(1)+3)/{\rm sinh}(2)$. This is a
volcano-type potential. That is, the coefficient of the
delta-function term is negative, and the $g^2$ term is positive at
$z=0$ and decreases to 
\be
V(z\rightarrow \infty)=g^2. \label{V}
\ee
The massless wavefunction solution is given by
\be \psi=N\sqrt{\frac{{\rm cosh}(1+g|z|)}{{\rm sinh}^3 (1+g|z|)}},
\label{massless} \ee
where $N$ is the normalization constant. Since this wavefunction
is square normalizable, it corresponds to a localized massless
graviton state. (\ref{V}) indicates that there is a mass gap of $M_{\rm
gap}=g$ separating the localized massless state from the massive modes
which propagate in the extra dimension.

Notice that the wave equation (\ref{wave}), and therefore the graviton
spectrum, is independent of $\gamma$. However, from (\ref{background}) we 
see that the speed of gravitons in the bulk depends on $\gamma$: 
\be
v(z)= \frac{{\rm cosh}(1+g|z|)}{\sqrt{1+\gamma^2 {\rm sinh}^2 (1+g|z|)}}.
\ee
As shown in Figure 1, $v(z)$ increases indefinitely away from the
braneworld for $\gamma=0$. For $\gamma >0$, $v_{maximum}=1/\gamma$. 

\begin{figure}
   \epsfxsize=4.0in
   \centerline{\epsffile{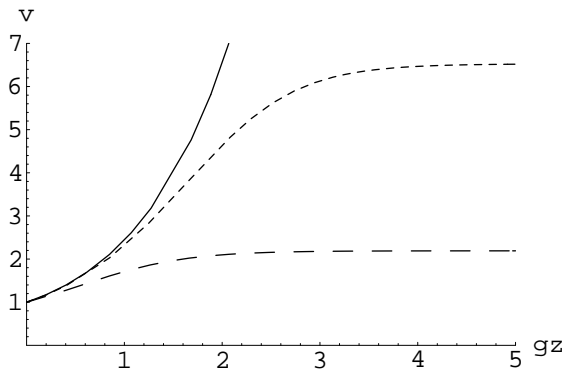}}
   \caption[FIG. \arabic{figure}.]{$v(z)/v(0)$ versus $gz$ for a
braneworld from a uniform disc of extremal D3-branes with
unspecified $k$. $v(z)$ increases indefinitely for $\gamma=0$ (regular
line). For $\gamma >0$, $v(z)$ asymptotes to a maximum of $1/\gamma$,
which is shown for $\gamma=.01$ (dotted line) and $.1$ (dashed line)} 
\end{figure}

Since $v(z)$ increases away from the braneworld, there is an apparent
violation of causality. That is, gravitational disturbances bend into
the bulk and arrive at a particular location on the braneworld earlier
than does the light from the same source, the latter of which
are restricted to the braneworld. Analogous to Fermat's Principle
for the propagation of light, gravitational waves will take the path of
least-time, given by the geodesics. 

\subsection{D3-brane with $k=1$ and the apparent speed of gravity}

The near-horizon region of a single D3-brane with $k=1$ is global $AdS_5
\times S^5$. This is the simplest supergravity solution that yields a
braneworld exhibiting an apparently ``superluminal" speed of gravity.
For this solution, one can exactly calculate the apparent average speed of
gravity between two points on the braneworld\footnote{The apparent
graviton speed is exactly calculable for the solution of general $\gamma$
from the previous section but we restrict ourselves to $\gamma=0$ for
simplicity.}. Thus, this example serves to illustrate characteristics which 
carry over to the more complicated braneworlds that exhibit apparently
``superluminal" speeds of gravity, which will be discussed in section 4. 

\begin{figure}
   \epsfxsize=4.0in
   \centerline{\epsffile{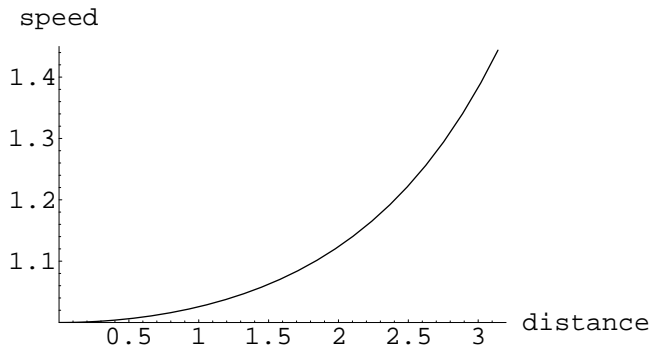}}
   \caption[FIG. \arabic{figure}.]{Apparent average graviton speed
$v_{\rm average}$ versus distance $gx_{\rm total}$ for a braneworld from
an extremal D3-brane with $k=1$.} 
\end{figure}

The geodesics can be deduced from the Lagrangian corresponding to the
background metric (\ref{background}),
\be
{\cal L}=\frac{1}{2}g_{MN}{\dot x^M}{\dot x^N}=\frac{1}{{\rm sinh}^2
(1+g|z|)}\Big( -{\rm cosh}^2 (1+g|z|)\, {\dot t}^2+ {\dot x}_i^2+ {\dot 
z}^2 \Big), \label{L} 
\ee
where the dot represents differentiation with respect to an affine
parameter. The equations of motion for $t$ and $x_i$ yield
\be
{\dot t}=E\, {\rm tanh}^2 (1+g|z|),\quad {\dot x}_i= p_i\, {\rm sinh}^2
(1+g|z|),
\ee
where $E$ and $p_i$ are constants of integration. Together with the
requirement that the Lagrangian (\ref{L}) vanishes for lightlike
geodesics, this yields
\be
{\dot z}^2+{\rm sinh}^4 (1+g|z|) \Big( p_i^2-\frac{E^2}{{\rm cosh}^2
(1+g|z|)} \Big) =0.
\ee
The turning point $z_T$ of gravitons in the bulk is at ${\dot
z}=0$: $E^2=p_i^2\, {\rm cosh}^2 (1+g|z_T|)$. 
$\tau$ divides out of the ratio ${\dot x}/{\dot z}$, and the remaining
expression can be integrated to obtain $x(z)$. The distance on the brane
after which a geodesic returns to the brane is given by
\be
x_{{\rm total}}=2\int_0^{z_T}\frac{dz}{\sqrt{\frac{{\rm
cosh}^2(1+g|z_T|)}{{\rm cosh}^2(1+g|z|)}-1}}.
\ee
Likewise, the corresponding time interval can be expressed as
\be
t_{{\rm total}}=2\int_0^{z_T}\frac{{\rm cosh}
(1+g|z_t|)\, dz}{{\rm cosh}^2 (1+g|z|) \sqrt{\frac{{\rm
cosh}^2(1+g|z_T|)}{{\rm cosh}^2(1+g|z|)}-1}}.
\ee
The apparent average speed with which gravitons propagate a given distance
along the braneworld is given by 
\be
v_{average}=\frac{x_{\rm total}}{t_{\rm total}}=
\frac{gx_{\rm total}}{\pi-2\, {\rm arctan}\Big( \sqrt{{\rm
sinh}^2(1) +\frac{{\rm cosh}^2(1)}{{\rm tan}^2(gx_{\rm total}/2)}\Big)}},
\ee
which is plotted in Figure 2.

\subsection{Nonextremal D3-branes}

\begin{figure}
   \epsfxsize=4.0in
   \centerline{\epsffile{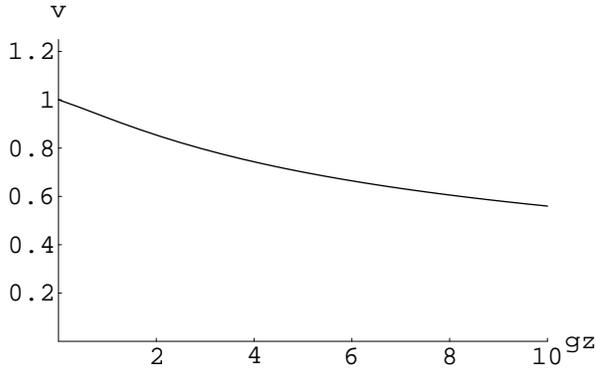}}
   \caption[FIG. \arabic{figure}.]{$v(z)/v(0)$ versus $gz$ for
a braneworld from a nonextremal D3-brane} 
\end{figure}

The case of $\mu \ne 0$ and vanishing $R_i$ corresponds to a nonextremal
D3-brane. From (\ref{coord}) we find that $r(z)$ can be written in terms 
of Jacobi polynomials. However, for our purposes it suffices to use a
less complicated coordinate\footnote{For typographical simplicity, we will 
drop the $^{\prime}$ in the remainder of this paper.}
\be
r=e+\frac{e}{1+g|z^{\prime}|}, \label{coordinate}
\ee
where $e\equiv (\mu/g^2)^{1/4}$. The event horizon is located where $f=0$,
at $r_h=e$. The graviton speed is given by
\be
v(z)=\sqrt{1-\Big( 1+\frac{1}{1+g|z|}\Big) ^{-4}}.
\ee

As can be seen from Figure 3, $v(z)$ decreases away from $z=0$.
Thus, gravitons do not appear to travel faster than the speed
of light. On the contrary, graviton geodesics bend {\sl towards}
the braneworld. A consequence of this is that the effective
four-dimensional graviton is a quasi-localized {\sl massive} state, whose
mass increases when further from extremality. A braneworld scenario with
such charactistics which is tractable with semi-analytical methods arises
from a non-extremal 5-brane in ten dimensions \cite{poritz, vazquez}. 

While moving away from extremality induces a mass in the gravitational
bound state from the braneworld viewpoint, from the holographic viewpoint it 
corresponds to giving a finite temperature to the dual gauge theory.

\section{Braneworlds from spinning D3-branes: phase transitions and
``superluminal" gravitons}

We now consider a nonextremal, spinning D3-brane. Above certain values of
angular momenta, the Hawking temperature vanishes and there is no longer
an event horizon hiding the singularity. There is evidence suggesting
that this point in the parameter space marks the transition from the 
deconfined high-density phase of the dual field theory to the Coulomb
phase at finite density \cite{evans}. For the case of one non-zero angular
momentum, this phase transition occurs at $\mu=0$. The braneworld scenario
arising from this point in parameter space was briefly reviewed in section
3.1.

\subsection{Phase transition for two angular momenta}

We now consider a nonextremal, spinning D3-brane with two angular momenta
with the value $R$. For $g^2 R_{{\rm transition}}^4=\mu$, there is no longer 
an event horizon, since $f=g^2(2R^2+r^2)$ is non-singular. In this case, 
(\ref{metric}) and (\ref{coord}) simplify to
\be
ds_5^2=\frac{(1+{\rm cosh}^2(1+g|z|))^{2/3}}{{\rm sinh}^2(1+g|z|)}
\Big( -\frac{{\rm cosh}^2(1+g|z|)}{(1+{\rm cosh}^2(1+g|z|))^2} dt^2
+dx_i^2+dz^2 \Big),
\ee
where
\be
r=\frac{\sqrt{2}R}{{\rm sinh}(1+g|z|)}.
\ee

The fluctuations of the five-dimensional graviton satisfy the equation for
a minimally-coupled scalar field given by (\ref{minimal}), which leads to
the wave equation (\ref{eom}). Therefore, there is a localized massless
graviton state separated from the massive modes by a mass gap of $M_{\rm
gap}=g$. In the previous section we mentioned that, for a static D3-brane,
going further from extremality by increasing $\mu$ serves to increase the
mass of the quasi-localized graviton. If we consider that $\mu$ is
then kept fixed while two angular momenta $R$ are increased from zero to
the phase transition value $R_{{\rm transition}}$, we find from the above
that the effect has been to render the localized graviton massless once
again. This is because, as $R$ is increased, the graviton speed does not
decrease as much in the bulk.

\subsection{Phase transition for three angular momenta}

\begin{figure}
   \epsfxsize=4.0in
   \centerline{\epsffile{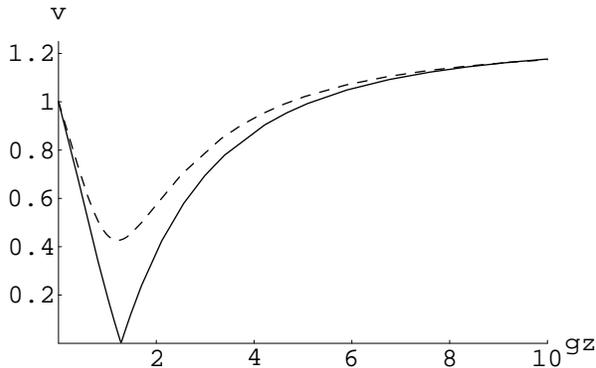}}
   \caption[FIG. \arabic{figure}.]{$v(z)/v(0)$ versus $gz$ for a
braneworld from a nonextremal spinning D3-brane with all three $R_i$
equal. $R=R_{\rm transition}$ corresponds to the graviton speed going to
zero at some point in the bulk and then rising greater than the
speed-of-light on the brane (regular line). For $R>R_{\rm transition}$,
the graviton speed reaches a non-zero minimum before rising greater than
the speed-of-light on the brane, which is shown for $g^2 R^4=.17\, \mu$ 
(dashed line).} 
\end{figure}

We will now consider the case of three equal $R_i=R$. For $R<R_{{\rm
transition}}$, where $g^2 R_{{\rm transition}}^4 = \frac{4}{27} \mu$,
there is an event horizon around the singularity. This corresponds to an
AdS-Reissner-Nordstr\"{o}m black hole from the five-dimensional
viewpoint. This has been interpreted as the high density and high
temperature deconfined phase of the dual {\cal N} gauge theory.

$R>R_{{\rm transition}}$ corresponds to a naked singularity.
These backgrounds have been used in previous studies of gravitational
Lorentz violations in braneworld scenarios \cite{dan,csaki,csaki2,csaki3}. 
There is evidence that suggests this to be dual to the Coulomb phase at 
finite density of the gauge theory \cite{evans}. 

From (\ref{coord}) we find that $r(z)$ can in general be written in
terms of Jacobi polynomials. However, for our purposes it suffices to use
the less complicated coordinate given by (\ref{coordinate}). The graviton
speed is then given by
\be
v(z)=\sqrt{1-\frac{\Big( 1+\frac{1}{1+g|z|}\Big) ^2}{\Big[
\Big( 1+\frac{1}{1+g|z|}\Big) ^2 +(R/e)^2 \Big]^3}}.
\ee
For $R<R_{{\rm transition}}$, $v(z)$ decreases away from $z=0$, as in the 
case of the static nonextremal D3-brane studied in section 3.3. At 
$R=R_{\rm transition}$, for which the bulk singularity becomes
naked, $v(z)$ goes to zero at some point in the bulk and then rises
to a speed which is greater than $v(0)$, the speed-of-light on the brane. 
For $R>R_{\rm transition}$, the graviton speed reaches a non-zero minimum
before rising greater than $v(0)$. This behavior is shown in Figure 4. 
As $R$ increases further, the change in $v(z)$ with respect to $z$ is 
lessened, as shown in Figure 5. Note that the minimum of $v(z)$ in the
bulk is too small to be seen in this plot.

\begin{figure}
   \epsfxsize=4.0in
   \centerline{\epsffile{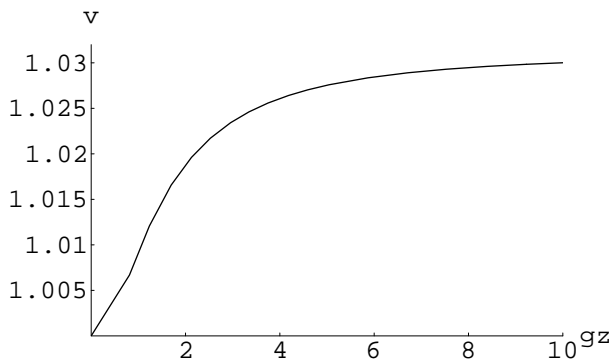}}
   \caption[FIG. \arabic{figure}.]{$v(z)/v(0)$ versus $gz$ for a
braneworld from a nonextremal spinning D3-brane with all three $R_i$ given
by $g^2 R^4 = 2.4\, \mu$.} 
\end{figure} 

While causality is not actually violated from the higher-dimensional
vantage point, the apparently ``superluminal" gravitons imply the presence
of a tachyon in the dual gauge theory. Indeed, at zero temperature
($R=R_{{\rm transition}}$) a spurious gauge field vev produces a negative
mass scalar term which destabilizes the moduli space of the theory
\footnote{The author thanks Paolo Creminelli for pointing out this
connection.} \cite{nick}. On the supergravity side, this corresponds to
radial motion of the spinning D3-branes.

One consequence of the $v(z)$ minimum, shown in Figure 4, is that graviton
geodesics close to the braneworld will actually bend {\sl towards} the
braneworld. This means that there is a quasi-localized {\sl massive}
state, as in the case of the static nonextremal D3-brane discussed in
section 3.4. As $R$ increases, the mass of the quasi-bound graviton
decreases.

As the energy scale increases, gravity on the braneworld probes less of
the bulk. That is, the geodesics connecting two points on the braneworld
do not have time to stretch as far into the bulk when the distance
between the points is decreased. Therefore, the minimum in $v(z)$ means
that there is a maximum energy for which gravity propagates at apparently
``superluminal" speeds. This is clearly desirable, since otherwise the
gravitational loops which transmit Lorentz violations would not be
suppressed at the energy scale where gravity becomes strongly interacting.

\subsection{Stability domains}

\begin{figure}
   \epsfxsize=4.0in
   \centerline{\epsffile{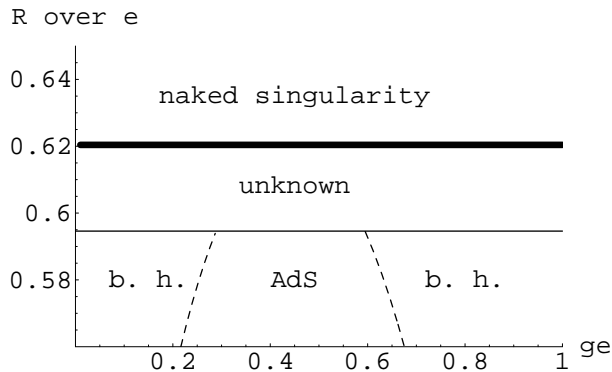}}
   \caption[FIG. \arabic{figure}.]{Stability domains for a
five-dimensional AdS-black hole with three equal charges $R$. 
Dashed lines mark the  Hawking-Page phase transition between a charged AdS
black hole (b.h.) and pure AdS. Above the regular line, the AdS black hole
is unstable. Above the bold line, there is a naked singularity. For
braneworld scenarios, this may be the most interesting region, since here
the graviton speed increases in the bulk.} 
\end{figure} 

We shall now consider the thermodynamical stability domains for a
spinning D3-brane 
\cite{cvet1,terning1,gubser,cai,russo,terning2,johnson1,cvet2,johnson2,harmark}. We
restrict ourselves to the case of three equal angular momenta $R$. There
is evidence that the grand-canonical ensemble, rather than the canonical
ensemble, is relevant when the D-brane world-volume is large. Therefore,
we shall work with the former. The Hawking-Page phase transition
(dashed lines in Figure 6) between a five-dimensional charged AdS black
hole and pure AdS is obtained by setting the Gibbs Euclidean action to
zero, as given in \cite{cvet1,cvet2}. This gives 
\be
(R/e)^2=(ge)^{2/3}-(ge)^2, 
\ee
where $e$ is defined by $g^2 e^4 \equiv \mu$. From the braneworld
perspective, the black hole region of parameter space corresponds to a
quasi-localized massive graviton, whereas the pure AdS region corresponds
to the original Randall-Sundrum model with a localized massless graviton. 

Thermodynamic stability is formally equivalent to the sub-additivity of
the entropy function, which for D3-branes amounts to the condition
\cite{cvet1,cvet2}:
\be
2-3(R/r_h)^2+(R/r_h)^6 \ge 0.
\ee
Numerically, this yields $R/e \approx .595$ (regular line in Figure 6). It
has been conjectured that, above this threshold, the D3-branes split apart
into fragments which move out in the radial direction carrying away some
of the spin. The resulting geometry may be neither stationary nor static
\cite{gubser}. One can conjecture that the corresponding braneworld has a
mass gap separating a quasi-localized massive graviton from a discrete
graviton spectrum.

The region above $R=R_{{\rm transition}}$ (bold line in Figure
6) corresponds to a naked singularity. The singularity may ultimately be 
smoothed out by the true short-distance theory of gravity. Thus, without
understanding the physics of the singularity, we cannot determine yet
whether it significantly affects the interactions of the four-dimensional
modes. Nevertheless, this may be the most interesting region from the
braneworld perspective, since it seems to exhibit apparently
``superluminal" gravitons.

However, it should be noted that contributions from higher-derivative
gravity \cite{noj1,noj2}, as well as logarithmic corrections in black
hole thermodynamics \cite{noj3}, may have a significant effect on the
previously-mentioned phase structure.

\section{Discussion}

We have considered various types of braneworlds which may arise from a
D3-brane solution. Gravitons which appear to travel faster than the
speed-of-light result if the D3-brane is spatially spherical. We
calculated exactly the apparent graviton speed for this case. For a flat
D3-brane, ``superluminal" gravitons only result if the D3-brane is both 
nonextremal and has three nonzero spins that are above a critical value,
for which case there is a naked singularity in the bulk. It has been
conjectured, from the holographic viewpoint, that the naked singularity
marks the transition to the finite-density Coulomb phase of the dual gauge
theory. Lastly, we discussed how the thermodynamic stability domains of
the spinning D3-brane affect the physics of braneworld scenarios.

\section*{Acknowledgments}

I am very grateful to Malcolm Fairbairn for extensive discussions
throughout the course of this work. Research is supported in full by the
Francqui Foundation (Belgium), the Actions de Recherche Concert{\'e}es of
the Direction de la Recherche Scientifique - Communaut\'e Francaise de
Belgique, IISN-Belgium (convention 4.4505.86).

\end{document}